\documentclass[10pt,draft]{dis03}
\usepackage{epsf,amsmath}

\def\sint{$\sin^2 \theta_W~$}

\def\sbar{$\bar{s}$}

\begin{document}

\title{STRANGE SEA ASYMMETRY FROM\\
GLOBAL QCD FITS}

\author{Benjamin Portheault \\
Laboratoire de l'Acc\'el\'erateur Lin\'eaire \\
IN2P3-CNRS et Universit\'e de Paris-Sud, BP 34, F-91898 Orsay Cedex\\
portheau@lal.in2p3.fr }

\maketitle

\begin{abstract}
\noindent We present a preliminary account 
of a new global QCD analysis of DIS data, including recent 
$\nu, \bar \nu$ DIS measurements. 
The model-independent cross section 
reanalysis by CCFR allows 
a new determination of the strange sea asymmetry, 
whose first moment is found to be small.
The impact on the NuTeV measurement of $\sin^2 \theta_W~$ is discussed.
\end{abstract}

\section{QCD analysis and parton distributions}

This analysis, which is an update of \cite{BPZ}, aims at extracting 
flavour-separated parton distributions from a global NLO QCD analysis of 
inclusive DIS and Drell Yan cross sections.
The data set used \cite{refBPZ} ranges from the cornerstone DIS data of BCDMS 
and NMC, including  recent data form HERA 
(neutral and charged current from H1 and Zeus), 
to the fixed-target Drell-Yan measurements of E866, E605 and E772. 
As for neutrino DIS, CCFR  cross section measurements - 
now available \cite{Yang} as a model independent result - 
together with the CDHSW data provide key constraints for the strange sea 
determination.

The parton distributions 
$g$, $u_v$, $d_v$, $\bar{u}+\bar{d}$, $\bar{d}-\bar{u}$
 and $s$, \sbar~ are parametrised and evolved in the 
NLO Fixed-Flavour Scheme. The strange sea is not constrained to have 
the same shape as the light sea, and $s\neq\bar{s}$ is allowed 
with the condition $\int_0^1 (s-\bar{s})\mathrm{d}x=0$ that ensures 
zero net strangeness. Details of the fit can be found in \cite{BPZ} 
and in a forthcoming paper.

\section{Extraction of the strange sea asymmetry}

Our focus in this communication will be on the strange 
sea asymmetry $s - \bar s$.
The requirement $s=$\sbar~ is not dictated by any symmetry of QCD, 
and qualitative models predict 
a significant asymmetry in the high $x$ region  arising from 
long-living higher Fock states containing intrinsic 
$s \bar s$ pairs \cite{brodsky}. 

\begin{figure}[!thb]
\vspace*{4.0cm}
\begin{center}
\includegraphics{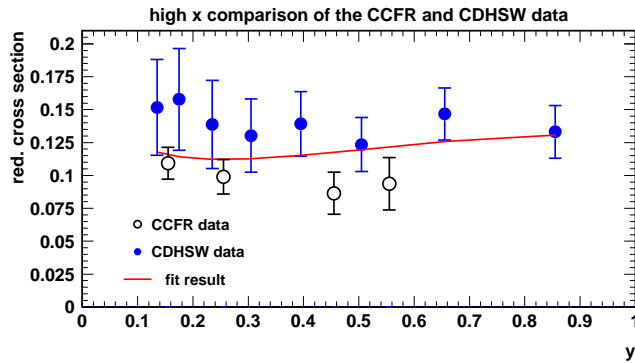}
\caption[*]{Comparison of the reduced 
neutrino cross sections of CCFR and CDHSW for $x=0.65$ and $E_\nu=110$ GeV. 
The normalisation shifts  determined by the fit are applied.}
\label{myfig}
\end{center}
\end{figure}

Experimental constraints on $s - \bar s$ come from charged current 
$\nu$ and $\bar{\nu}$ cross sections: the quantity 
\begin{equation}
\frac{\mathrm{d}^2\sigma^{\nu N}}{\mathrm{d}x\mathrm{d}Q^2}-
\frac{\mathrm{d}^2\sigma^{\bar{\nu} N}}{\mathrm{d}x\mathrm{d}Q^2}
\propto xs-x\bar{s}+\left[ 1-(1-y^2) \right](xu_v+xd_v)
\end{equation}
\noindent
valid at LO for an isoscalar target, exhibits the sensitivity 
to the strange sea asymmetry (valence distributions are well constrained 
by other data). Both CCFR and CDHSW provide high $x$ data: the latter
tend to be higher for neutrinos (see figure \ref{myfig}), 
whereas the two series of data are in agreement for antineutrinos. 
The NuTeV Collaboration has recently released new data 
on $\nu, \bar \nu$ DIS with dimuon production \cite{dimuons}. 
These data are not statistically significant for $x>0.5$ 
but constrain the strange sea at small $x$ and affect 
indirectly the large-$x$ region because of the strange number sum rule
$\int (s - \bar s) \mathrm{d} x = 0$. Unfortunately, NuTeV 
data cannot be included 
in a global cross section fit in the form they are published.  

\begin{figure}[!thb]
\vspace*{5.0cm}
\begin{center}
\includegraphics{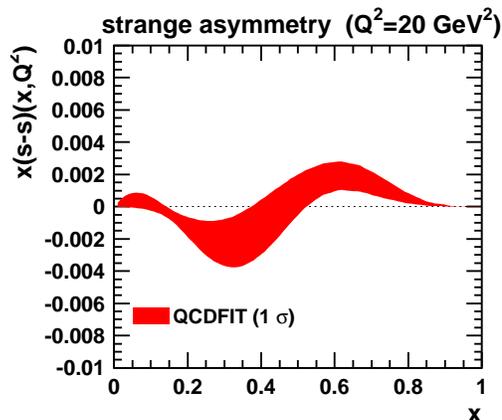}
\caption[*]{The strange sea asymmetry  (one sigma error band) as obtained with the global fit including all data.}
\label{fig2}
\end{center}
\end{figure}

\noindent
With both CDHSW and CCFR data sets we found that the strange 
sea asymmetry is $\int x(s-\bar{s})\mathrm{d}x=(1.8\pm3.8)\times10^{-4}$ (see figure \ref{fig2}). 
Removing the CCFR data leads to a larger asymmetry, 
$\int x(s-\bar{s})\mathrm{d}x=(1.8\pm0.5)\times10^{-3}$, 
as already shown by our previous results \cite{BPZ}. 
It should be noticed that the dominant systematic uncertainty 
(flux normalisation) for CCFR neutrino data is fitted to 
$-5.2\sigma$ (the data are scaled up by 4\%), irrespectively  
of the assumptions on the strange sea.  
On the other hand, there is the well known 
problem  of the CDHSW  $y$ shape for $x<0.1$ which do not follow 
the QCD prediction. 
A reasonable attitude is to take both data sets 
into account in the global analysis keeping in mind that 
one of them or both may be affected by uncontrolled experimental effects.
A recent study by CTEQ \cite{tung}, which includes the NuTeV dimuon data, 
shows that these data drive a bump of $s(x)- \bar s(x)$ 
in the medium-large $x$ region, in qualitative agreement 
with the findings of \cite{BPZ}.

\section{Impact on the NuTeV \sint measurement}

The NuTeV experiment uses a 
fit to the measured ratios of neutral current to 
charged current cross sections $R^{\nu (\bar \nu)}
= \sigma^{\nu (\bar \nu)}_{NC}/\sigma^{\nu (\bar \nu)}_{CC}$
to extract \sint. 
They obtain $\sin^2 \theta_W=0.2277\pm0.0016$ \cite{nutev} 
which is $3.1 \sigma$ away from the fit result 
by the LEP EWWG of $\sin^2 \theta_W=0.22272\pm0.00036$ \cite{LEPEWWG}.
A related quantity is the Paschos-Wolfenstein ratio
$R^-=(\sigma^{\nu}_{NC}-\sigma^{\bar{\nu}}_{NC})/
(\sigma^{\nu}_{CC}-\sigma^{\bar{\nu}}_{CC})$,  
given by $R^-=1/2-\sin^2 \theta_W$ 
at LO for an isoscalar target. 
Discussion of the NuTeV determination 
of \sint in the context of the present work is 
relevant since $R^-$ must be corrected if there is a 
fractional neutron excess $\delta N$ or if  
$s\neq\bar{s}$. In this case 
$R^-$ reads \cite{OldNew,Kulagin}

{\setlength\arraycolsep{2pt}
\begin{eqnarray}
R^{-} & = & \frac{1}{2}-\sin^2 \theta_{W} 
      -\left(\delta N 
\frac{\int x(u_v-d_v)\mathrm{d}x}{\int x(u_v+d_v)\mathrm{d}x} 
+ \frac{\int x(s-\bar{s})\mathrm{d}x}{\int x(u_v+d_v)\mathrm{d}x} \right) 
\nonumber \\
      &   & \times\left[ 1-\frac{7}{3}\sin^2 \theta_W 
+ \frac{8 \alpha_s}{9\pi}\left(\frac{1}{2}-\sin^2 \theta_W\right)\right] 
\nonumber \\
      & \equiv &  \frac{1}{2} - \sin^2 \theta_{W} +\delta R^{-} \label{deltar}
\end{eqnarray}
}
\noindent
The correction $\delta R^{-}$ as well as a properly propagated error 
can be computed using the parton distributions of this analysis. 
With all data sets one obtains  $\delta R^{-}=-0.0107\pm0.0005$,
without the CCFR data set  the significant strange sea asymmetry 
leads to the value $\delta R^{-}=-0.0135\pm0.0008$. 
whereas the NuTeV collaboration reports the value \cite{kevin}  
$\delta R^{-}=-0.0080\pm0.00005$. 
Taking the difference between these corrections one can roughly estimate 
the corresponding shift of \sint: without  the CCFR data 
the strange sea asymmetry has the required magnitude 
to reduce $\sin^2 \theta_W$ to $0.2222\pm0.0018$, 
and with all data included $\sin^2 \theta_W=0.2249\pm0.0017$, 
which is now $1.35 \sigma$ away from the Standard Model fit value. 
Here the reduction is due to the neutron excess correction 
which is larger in our case, but approximately half of this discrepancy 
can be understood by taking into account the experimental cuts 
and cross talk between NC and CC  using the model described in \cite{sven}. 
However for a realistic estimate a full MC study is required. 
Another relevant point is the parton distribution uncertainty,
which is found to be one order of magnitude larger than 
the reported NuTeV value.

Since the \cite{kevin} the NuTeV collaboration have re-evaluated 
this parton distribution error adjusting the value\footnote{These numbers replace the mistyped values contained in  the DIS03 proceeding.} $0.00005$ to $0.0003$ \cite{zeuthen}. 
Note that the error in the evaluation have been independently found 
and originally reported to the NuTeV collaboration  
by S. Alekhin and S. Kulagin.

\section*{Acknowledgements} I would like to thank  S. Kulagin, K. McFarland, G. P. Zeller for useful discussions on the NuTeV analysis, and S. Alekhin, W.K. Tung for very interesting discussions at the DIS workshop.

\end{document}